\documentclass[preprint,showpacs,preprintnumbers,amsmath,amssymb,aps]{revtex4}
\usepackage{epsfig}
\usepackage{psfrag}
\usepackage{graphicx}
\usepackage{color,graphics}

\begin{document}

\title{Neutrino oscillations in a turbulent plasma}

\author{J.T. Mendon\c{c}a}

\email{titomend@ist.utl.pt}

\affiliation{Instituto de F\'isica, Universidade de S\~ao Paulo, S\~ao Paulo SP, CEP 05508-090 Brasil
\\ IPFN, Instituto Superior T\'ecnico, 1049-001 Lisboa, Portugal}

\author{F. Haas}
\affiliation{Departamento de F\'isica, Universidade Federal do Paran\'a, Curitiba PR, CEP 81531-990, Brasil}

\begin{abstract}
A new model for the joint neutrino flavor and plasma oscillations is introduced, in terms of the 
dynamics of the neutrino flavor polarization vector in a plasma background. Fundamental solutions 
are found for both time-invariant and time-dependent media, considering slow and 
fast variations of the electron plasma density. The model is shown to be described by a 
generalized Hamiltonian formalism. In the case of a broad spectrum of electron plasma waves, 
a statistical approach indicate the 
shift of both equilibrium value and frequency oscillation of flavor coherence, due to the existence of a
turbulent plasma background.
\end{abstract}


\maketitle

\section{Introduction}

Neutrinos are elusive particles which interact very weakly with matter, but play an important and sometimes a decisive role in astrophysical phenomena \cite{raffelt}. In recent years, the success of experimental neutrino physics has been spectacular, providing detailed information on the stellar matter processes, as well as on their intrinsic properties in vacuum \cite{duan}. The existence of a finite neutrino mass leads to the occurrence of neutrino flavor oscillations in vacuum, as documented by experiments and observations. 

Neutrino interactions with a dense plasma show that a significant amount of energy transfer between neutrino beams and plasma waves can take place over distances, thus suggesting that such a mechanism could be crucial for the formation of an outgoing shock in type II supernovae \cite{bobrev}. Such a coupling results from the existence of an induced neutrino charge \cite{semikoz, nieves, serbeto}, leading to collective kinetic instabilities, which are mediated by neutrino Landau damping \cite{prlneu}. Neutrino plasma interactions could also lead to the emission of electron-positron pairs \cite{luciana}, and to the excitation of quasi-static magnetic fields \cite{padma}. On the other hand, the interaction with matter can also significantly change the neutrino flavor oscillations, and lead to a resonant coupling between neutrino flavor states, known as the MSW effect \cite{bethe,wolf,mikhey}.

Neutrino plasma instabilities and neutrino flavor oscillations are usually considered as two distinct phenomena, and have been explored independently by two different scientific communities. Here we try to establish the bridge between these two communities, by considering the influence of plasma instabilities and plasma turbulence on the neutrino flavor oscillations. Our approach is also motivated by recent studies of neutrino behavior in a turbulent background \cite{volpe}. 

In the present work we consider the influence of plasma oscillations on the evolution of the neutrino flavor polarization vector. The influence of a space or time varying medium on the neutrino flavor content has been considered by many authors, but only a few \cite{schafer,krastev} consider sine-variations (in time) of the electron density, as we do here. In contrast with this previous work, which is mainly concerned with numerical solutions of the neutrino oscillation equations, we use here a  WKBJ-like method, which allows us to derive simple analytical solutions in which the role of higher harmonics becomes apparent. In addition, several other papers in this area concern the influence of a stochastic background medium (see for instance \cite{torrent,benatti}), or discuss general time-dependent media \cite{hollenberg}. 

This work is organized as follows. In Section II we define the flavor polarization vector, set the basic equations of our model and derive their solutions for a plasma in steady-state. In Section III we consider flavor oscillations in a time-varying medium, and discuss the cases of very slow and very fast plasma oscillations. In Section IV we demonstrate that, even for an arbitrary temporal variation of the electron plasma density, the evolution of the polarization vector  can be described by a generalized Hamiltonian formalism. In Section V we consider the case of a broad spectrum of electron plasma oscillations. Finally, in Section VI, we state our conclusions.

\section{Basic description}

In order to allow a direct comparison with the time-dependent situation, in this Section we review the basic neutrino oscillations in the autonomous case. 
It is well known that neutrino mass eigenstates $\left| \nu_j \right>$, with $j = 1, 2, 3$, differ from the neutrino flavor eigenstates $\left| \nu_\alpha \right>$, with $\alpha = e, \mu, \tau$, which are identified in weak interaction processes. This leads to a flavor oscillation process, first suggested by Pontecorvo \cite{ponte}, and then put on a more solid basis by Maki et al. \cite{maki}. These different eigenstates are related by a transformation matrix, according to
$\left| \nu_\alpha \right> = \sum_j U_{\alpha j} \left| \nu_i \right>$, where the neutrino mixing matrix $U_{\alpha j}$, known as the PMNS (Pontecorvo-Maki-Nakagawa-Sakata) matrix, contains three mixing angles and one CP violating phase.

Here we concentrate on a simplified two-flavor model, with $i =1, 2$ and $\alpha = e, \mu$. The restriction to two flavors is very common in the literature, because it allows to derive explicit analytic results, which are very important for a qualitative analysis.
The two flavor states, $\left| \nu_e \right>$ and $\left| \nu_\mu \right>$, can then be seen as a linear combination of  the two mass eigenstates $\left| \nu_1 \right>$ and $\left| \nu_2 \right>$, as defined by the transformation
\begin{equation}
\left[ \begin{array}{c} \nu_e \\ \nu_\mu \end{array} \right] = {\bf U} (\theta_0) \cdot \left[ \begin{array}{c} \nu_1 \\ \nu_2 \end{array} \right] 
\; , \quad {\bf U} (\theta_0) = \left[ \begin{array}{c} \cos \theta_0 \quad \sin \theta_0 \\  -\sin \theta_0 \quad \cos \theta_0 \end{array}  \right] ,
\label{2.1} \end{equation}
where $\theta_0$ is the relevant mixing angle.The temporal evolution of the mass eigenstates is obviously of the form $\left| \nu_j (t)\right> = \left| \nu_j (0) \right> \exp ( - i E_j t)$, for $j =1, 2$, and energies $E_j = (| {\bf p}_j |^2 + m_j^2)^{1/2}$, with $\hbar = 1, c = 1$. Because of the smallness of the masses $m_j$, we can use the approximation $E_j = p_j + m_j^2 / 2 p_j$. 
As for the temporal evolution of the flavor eigenstates, we can use Eq. (\ref{2.1}). Alternatively, we can consider the evolution of the density matrix $\rho$, with elements $\rho_{a b} = \psi_b^* \psi_a$, with $a, b = (e, \mu)$, where $\psi_a$ are the neutrino flavor wave functions. We can also define a three-dimensional flavor polarization vector ${\bf P}$, such that the density matrix can be defined as
\begin{equation}
\rho = \frac{N_0}{2} ( 1 + {\bf P} \cdot {\bf \sigma} ) ,
\label{2.2} \end{equation}
where ${\bf \sigma} \equiv (\sigma_1, \sigma_2, \sigma_3)$, and $\sigma_j$ for $j = 1,2, 3$ are the Pauli matrices. Here we have introduced the total number of neutrinos $N_0 = N_e + N_\mu$, as the sum of the the flavor populations. Alternatively, the single particle normalization condition, $N_0 = 1$ could also be used. It can then be shown \cite{raffelt} from (\ref{2.1}) that ${\bf P}$ evolves in time according to 
\begin{equation}
\frac{d {\bf P}}{d t} = \omega_0 \left( {\bf B} \times  {\bf P} \right) \; , \quad {\bf B} = \left[ \begin{array}{c} \sin 2 \theta_0 \\ 0 \\ \cos 2 \theta_0 \end{array} \right] .
\label{2.3} \end{equation}
Here we have introduced the characteristic oscillation frequency $\omega_0 = \Delta m^2 / 2 E$, where $\Delta m^2 = m_2^2 - m_1^2$ is the square mass difference and $E$ the fixed energy associated to the neutrino Dirac spinor. This is formally identical to the spin precession in a magnetic field, where ${\bf P}$ plays the role of a fictitious spin vector, and ${\bf B}$ is a fictitious magnetic field. In the presence of a background plasma, this evolution equation becomes
\begin{equation}
\frac{d {\bf P}}{d t} = \left( \omega_0  {\bf B} - \sqrt{2} G_F n_e {\bf L} \right)  \times  {\bf P} , 
\label{2.4} \end{equation}
where $G_F$ is the Fermi constant, $n_e$ is the electron plasma density and ${\bf L} \equiv (0, 0, 1)$. This can be rewritten in a form identical to the vacuum equation (\ref{2.3}), by introducing a new oscillating frequency $\omega$ and a new fictitious magnetic field ${\bf H}$, such that $\omega {\bf H} = \omega_0 {\bf B} - \sqrt{2} G_F n_e {\bf L}$. We then get
\begin{equation}
\frac{d {\bf P}}{d t} = \omega \left( {\bf H} \times  {\bf P} \right) \; , \quad {\bf H} = \left[ \begin{array}{c} \sin 2 \theta \\ 0 \\ \cos 2 \theta \end{array} \right] ,
\label{2.5} \end{equation}
where the new frequency $\omega$ and the new angle $\theta$ are determined by
\begin{equation}
\omega = \omega_0 \frac{\sin 2 \theta_0}{\sin 2 \theta} \; , \quad  \tan 2 \theta = \frac{\sin 2 \theta_0}{\cos 2 \theta_0 - \xi} \, ,
\label{2.6} \end{equation}
where the parameter $\xi = \sqrt{2} G_F n_e / \omega_0$ describes the neutrino plasma coupling. In explicit form, we have the coupled evolution equations for the three components of ${\bf P}$ as
\begin{equation}
\frac{d P_2}{d t} = \omega P_1 \cos 2 \theta  -\omega P_3 \sin 2 \theta ,
\label{2.7} \end{equation}
and
\begin{equation}
\frac{d P_1}{d t} = - \omega P_2 \cos 2 \theta  \; , \quad \frac{d P_3}{d t} = \omega P_2 \sin 2 \theta .
\label{2.7b} \end{equation}
From where we get  $d^2 P_2 / d t^2 = - \omega^2 P_2$, with the general solution
\begin{equation}
P_2 (t) = A \exp (- i \omega t) + B \exp (+ i \omega t) .
\label{2.8} \end{equation}
where $A$ and $B$ are integration constants. Replacing this in Eq. (\ref{2.7b}), we can also solve for the other two components, as
\begin{equation}
P_1 (t) = - i \cos 2 \theta \left[ A \exp (- i \omega t) - B \exp (+ i \omega t) \right] +P_{10} ,
\end{equation}
and 
\begin{equation}
P_3 (t) =  i \sin 2 \theta \left[ A \exp (- i \omega t) - B \exp (+ i \omega t) \right] +P_{30} ,
\label{p}
\end{equation}
where $P_{10}$ and $P_{30}$ are two additional constants. Since Eqs. (\ref{2.7}) and (\ref{2.7b}) constitute a system of three first-order ordinary differential equations, the general solution contain only three integration constants, hence not all $A, B, P_{10}$ and $P_{30}$ are independent. Indeed, by direct substitution it can be verified that 
\begin{equation}
P_{10} = C \sin 2 \theta \,, \quad P_{30} = C \cos 2 \theta \,,
\label{c}
\end{equation}
in terms of a single integration constant $C$ so that $A, B$ and $C$ are the independent arbitrary constants in the general solution.

Let us now relate these results with the neutrino density matrix. According to (\ref{2.2}), we have
\begin{equation}
\rho_{11} = \frac{N_0}{2} (1 + P_3) \; , \quad \rho_{22} = \frac{N_0}{2} (1 - P_3) \; , \quad \rho_{12} = \rho_{21}^* = \frac{N_0}{2} (P_1 - i P_2)  .
\label{2.10} \end{equation}
 In passing, notice that since $\rho$ is hermitian one need all components of the polarization vector to be real.  Moreover, in the weak interaction basis we can identify the diagonal matrix elements with the neutrino flavor populations, as $\rho_{11} = N_e$ and $\rho_{22} = N_\mu$, which only depend on $P_3 = (N_e - N_\mu) / N_0$. As for the coherences $\rho_{12}$ and $\rho_{21}$, they are determined by $P_1$ and $P_2$. The evolution equations for the neutrino populations are then given by
 \begin{equation}
 \frac{d N_e}{d t} = - \frac{d N_\mu}{d t} = \frac{N_0 \,\omega P_2}{2} \sin2 \theta .
 \label{2.11} \end{equation}

One needs $P_3$ real, implying $A = B^{*}, C = C^{*}$ in Eqs. (\ref{p}) and (\ref{c}). Therefore one has 
\begin{equation}
A = B^{*} = \frac{\alpha}{2}\,e^{i\beta} \,,
\end{equation}
where $\alpha, \beta$ are real numbers. Finally the solution can be expressed as
\begin{eqnarray}
P_3 &=& C \cos 2 \theta + \alpha \sin(\omega t - \beta) \sin 2 \theta \,, \nonumber \\
P_1 &=& C \sin 2 \theta - \alpha \sin(\omega t - \beta) \cos 2 \theta \,,\\
P_2 &=& \alpha \cos(\omega t - \beta) \,, \nonumber 
\end{eqnarray}
uniquely involving real integration constants $\alpha, \beta, C$. Note that physically acceptable solutions satisfy $|P_{3}| \leq 1$ for all time, otherwise negative flavor populations will eventually appear.

The constants of integration in the above solution have to satisfy the initial conditions. For instance, as a particularly important example, let us consider the case of an electron neutrino beam created at $t = 0$, as determined by the initial conditions
\begin{equation}
N_e (0 ) = N_0 \; , \quad N_\mu (0) = 0 \; , \quad P_3 (0) = 1.
\end{equation}
In this case, it is convenient to use the constants
\begin{equation}
\alpha = \sin 2 \theta \; , \quad \beta = - \frac{\pi}{2} \; , \quad C = \cos 2 \theta \, .
\end{equation}
One then finds 
\begin{eqnarray}
P_1 (t) &=&\cos 2 \theta \sin 2 \theta  \, [1 - \cos (\omega t)] \,, \nonumber \\
P_2 (t) &=& - \sin 2 \theta \sin (\omega t) \,, \label{x} \\
P_3 (t) &=& 1 - \sin^2 2 \theta \, [1 - \cos (\omega t) ]  \,, \nonumber 
\end{eqnarray}
This corresponds to neutrino flavor populations evolving as
\begin{equation}
N_e (t) = N_0 - \frac{N_0}{2} \sin^2 2 \theta \, [1- \cos (\omega t)]  \,, \quad 
N_\mu (t) =  \frac{N_0}{2} \sin^2 2 \theta \, [1- \cos (\omega t)]  \,.
\end{equation}
This coincides with the probability for an electron neutrino to become a muon neutrino, $P (\nu_e \rightarrow \nu_\mu, t) \equiv N_\mu (t) / N_0$, as known from direct calculations. The conservation of total number of neutrinos is obviously satisfied.
 
\section{Influence of plasma oscillations}

The basic solutions described above are modified in the presence of plasma density perturbations. In this case, a neutrino beam sees a time varying background electron density, and the quantities $\omega$ and $\theta$ become functions of (space and) time. Equations (\ref{2.7b}) are still valid, where the component $P_2$ is now determined by the  equation
\begin{equation}
\frac{d^2 P_2}{d t^2} + \omega^2  P_2 = \frac{d \ln \omega}{d t} \left[\frac{d P_2}{d t} + \omega \left(P_1 \sin 2 \theta + P_3 \cos 2 \theta\right) \tan 2 \theta\right] \,,
\label{3.1} \end{equation}
where we have used the first of Eqs. (\ref{2.6}). We can simplify the form of (\ref{3.1}) by means of a change of time parametrization, defining 
\begin{equation}
T = \int^{t}\omega(t')\, dt' \,,
\label{t} \end{equation}
which gives 
\begin{equation}
\frac{d^2 P_2}{dT^2} + P_2 = \frac{d \ln \omega}{d T} \left(P_1 \sin 2 \theta + P_3 \cos 2 \theta\right) \tan 2 \theta \,.
\label{h}
\end{equation}
In the case of very slow plasma processes, we can neglect the right hand side of (\ref{h}). This is valid for $| d \ln \omega / d t | \ll \omega$, and for a finite $\tan 2 \theta$. In this case, $d^2 P_{2}/dT^2 + P_2 \simeq 0$ and the solution (\ref{2.8}) can be replaced by the WKBJ (Wentzel-Kramers-Brillouin-Jeffreys) like form
\begin{equation}
P_{2}(t)  =  A \exp \left(- i \int^t \omega (t')  dt' \right) + B \left(+ i \int^t \omega (t')  dt' \right) .
\label{3.2} \end{equation}
The usual WKBJ expression contains the inverse square root of the frequency. However, in our case the rescaling of time completely eliminate the frequency, as far as the right-hand side of Eq. (\ref{h}) is negligible. This is because of the extra term $(\dot\omega/\omega) \dot{P}_2$ in Eq. (\ref{3.1}).
In passing, we note that assuming the reduction 
$P_1 \sin 2 \theta + P_3 \cos 2 \theta \equiv 0$,
one has exactly $d^{2} P_{1}/dT^2 + P_1 = d^{2} P_{2}/dT^2 + P_2 = 0$, which allows an exact solution regardless the form of the time-dependent frequency. 

Let us assume that we are in the presence of a plasma oscillation at some frequency $\omega' \ll \omega$. This could be the case of an ion acoustic mode, such that $\omega' = k' v_{ac}$, where $k'$ is the wavenumber and $v_{ac} = \sqrt{{\cal T} / m_p}$ is the ion acoustic velocity, with ${\cal T}$ the plasma temperature in energy units and $m_p$ the proton mass. We can then use
\begin{equation}
\omega (t) = \left< \omega \right> [1 + \epsilon \cos \omega' t] , 
\label{3.2b} \end{equation}
where $\left< \omega \right>$ is the frequency value in the absence of  plasma oscillations, and  $\epsilon$ is the amplitude of the electron density modulations.  More precisely, from Eq. (\ref{2.6}) we infer that time-dependent electron density modulations produce a changing frequency $\omega(t)$, modeled in a simple way through Eq. (\ref{3.2b}). 
The amplitude parameter $\epsilon$ can be simply interpreted as the ratio between the electron trapping frequency in the potential well of the electron plasma wave and the electron plasma frequency \cite{shukla}.

To avoid analytic difficulties, we will uniquely consider forward propagating waves, so that $0 \leq \epsilon < 1$. Notice that from Eq. (\ref{2.6}) one has 
\begin{equation}
\sin 2\theta = \frac{\omega_0 \sin 2\theta_0}{\left< \omega \right> (1 + \epsilon \cos \omega' t)} \,.
\end{equation}
The absolute value of the right hand side of the last equation should not exceed unity, otherwise ${\bf P}$ would became complex in a finite time, spoiling our physical interpretation. Hence the parameters should satisfy $\omega_{0} \sin 2\theta_0/\left[\left< \omega \right> (1-\epsilon)\right] \leq 1$. In addition, from Eq. (\ref{2.6}) some algebra shows that the equilibrium frequency value should exceed the vacuum value, $\left< \omega \right> > \omega_0$.
Finally, for the frequency (\ref{3.2b}) the slow time-dependence condition reads $\epsilon \omega'/\left< \omega \right > \ll 1$.

Replacing Eq. (\ref{3.2b}) in (\ref{3.2}), we get
\begin{equation}
P_2 (t) =  \sum_n J_n (\kappa) \left[A \exp \left(- i \left< \omega \right> t - i n \omega' t \right) + B \exp \left(i \left< \omega \right> t + i n \omega' t \right) \right] ,
\label{3.3} \end{equation}
where $J_n$ are Bessel functions with argument $\kappa = \epsilon\left< \omega \right>/\omega'$. The spectral broadening of the solution $P_2 (t)$ introduces a quantum decoherence 
which in a sense is equivalent to an energy broadening of the neutrino beam. This leads to a decrease of the amplitude of the neutrino flavor oscillations due to phase mixing, similar to Landau damping \cite{raffelt}. 

Repeating the procedure done for a steady-state medium and integrating Eq. (\ref{2.7b}) once to obtain $P_3$, a tedious analysis allows the solution to be expressed as 
\begin{eqnarray}
P_2 &=& \alpha \sum_n J_n (\kappa) \cos\left[\left(\left< \omega \right>  + n \omega'\right) t - \beta\right] \,,\\
P_3 &=& P_{3}(0) + \alpha \omega_0 \sin 2\theta_0 \sum_n \frac{J_n (\kappa)}{\left< \omega \right> + n \omega'} \sin\left[\left(\left< \omega \right>  + n \omega'\right) t - \beta\right]  \nonumber \\ 
&+& \alpha \omega_0 \sin 2\theta_0 \sin \beta \sum_n \frac{J_n (\kappa)}{\left< \omega \right> + n \omega'} \,, \label{p3}
\end{eqnarray}
in terms of real constants $\alpha, \beta$ and $P_{3}(0)$. We omit the complicated form of $P_1$. From Eq. (\ref{p3}) it can be proven that $P_{3}(t_*) - P_{3}(0) > 0$ after a finite time $t_*$ whose precise value depends on the initial conditions (the rough estimate $\epsilon \omega' t_* \simeq 1$ applies). Hence, starting from a pure electron neutrino beam ($P_{3}(0) = 1$) one would eventually get negative flavor populations, at least in the context of the slowly varying medium approximate solution. 
In specific cases, these difficulties can be avoided choosing $P_{3}(0) < 1$,  associated e.g. to a mixed state neutrino beam.

The spectral broadening induces beating in the oscillations of $P_3$. Consider, for instance, the parameters $\alpha = \sin 2\theta_0, \beta = - \pi/2$, so that $P_{2}(0) = 0$ and 
\begin{eqnarray}
P_2 &=& - \sin 2\theta_0 \sum_n J_n (\kappa) \sin\left[\left(\left< \omega \right>  + n \omega'\right) t \right] \,, \label{p2n} \\
P_3 &=& P_{3}(0) - \omega_0 \sin^{2}2\theta_0\sum_n \frac{J_n (\kappa)}{\left< \omega \right> + n \omega'} \left[1 - \cos\left(\left< \omega \right> + n \omega'\right)t \right] \,. \label{p3n}
\end{eqnarray}
When $\epsilon = 0$ and $P_{3}(0) = 1$, we immediately regain the solution (\ref{x}) valid for steady media, due to $J_{n}(0) = \delta_{n0}$. 

Figures \ref{f1} and \ref{f2} below show a representative case with $\omega_0 = 1.0, \left< \omega \right > = 1.1, \omega' = 0.1, \sin^{2} 2\theta_0 = 0.15, \epsilon = 0.1$ and $P_{3}(0) = 0.98$. In this situation the slow time variation condition $\epsilon \omega' \ll \left< \omega \right>$ is fairly well satisfied and one can employ Eqs. (\ref{p2n}) and (\ref{p3n}).

\begin{figure}
\centering
\includegraphics[width=8.5cm]{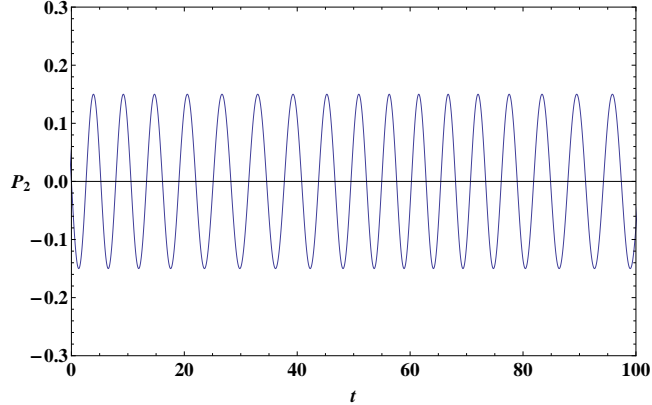}
\caption{{\sl Oscillations of the coherence variable $P_2$ from the approximate solution (\ref{p2n}) with parameters $\omega_0 = 1.0, \left< \omega \right > = 1.1, \omega' = 0.1, \sin^{2} 2\theta_0 = 0.15$ and $\epsilon = 0.1$.}}
\label{f1} \end{figure}

\begin{figure}
\centering
\includegraphics[width=8.5cm]{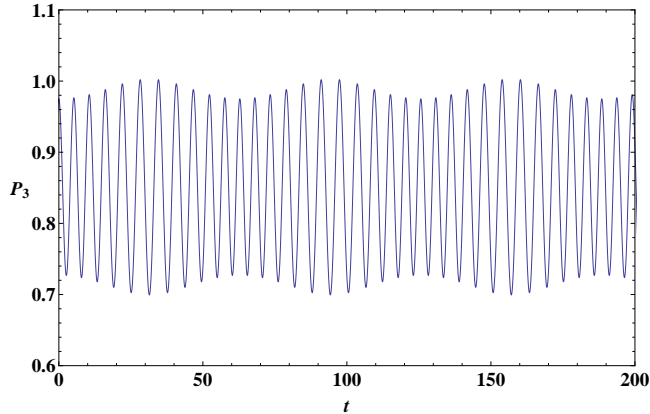}
\caption{{\sl Oscillations of the variable $P_3$ from Eq. (\ref{p3n}) using the same parameters of Fig. \ref{f1} together with $P_{3}(0) = 0.98$.}}
\label{f2} \end{figure}

Figures \ref{f3} and \ref{f4} below shows more complicated oscillations for strongly nonlinear plasma waves, with the same parameters of Figs. \ref{f1} and \ref{f2}, except that now $\epsilon = 0.6$ and $P_{3}(0) = 0.87$. The slow time varying assumption is still fairly well satisfied.  

\begin{figure}
\centering
\includegraphics[width=8.5cm]{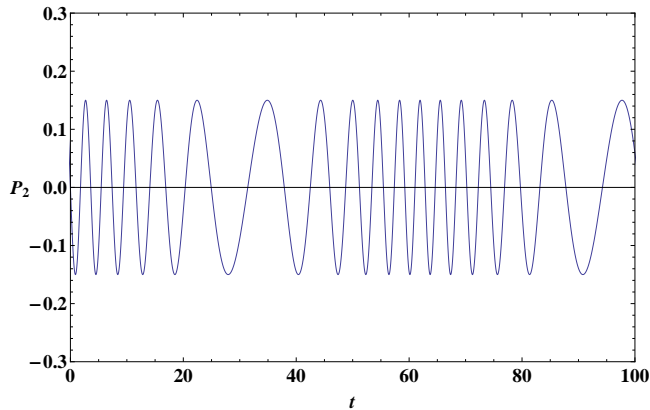}
\caption{{\sl Oscillations of the coherence variable $P_2$ from the approximate solution (\ref{p2n}) with the same parameters of Figs. \ref{f1} and \ref{f2}, except that now $\epsilon = 0.6$.}}
\label{f3}\end{figure}

\begin{figure}
\centering
\includegraphics[width=8.5cm]{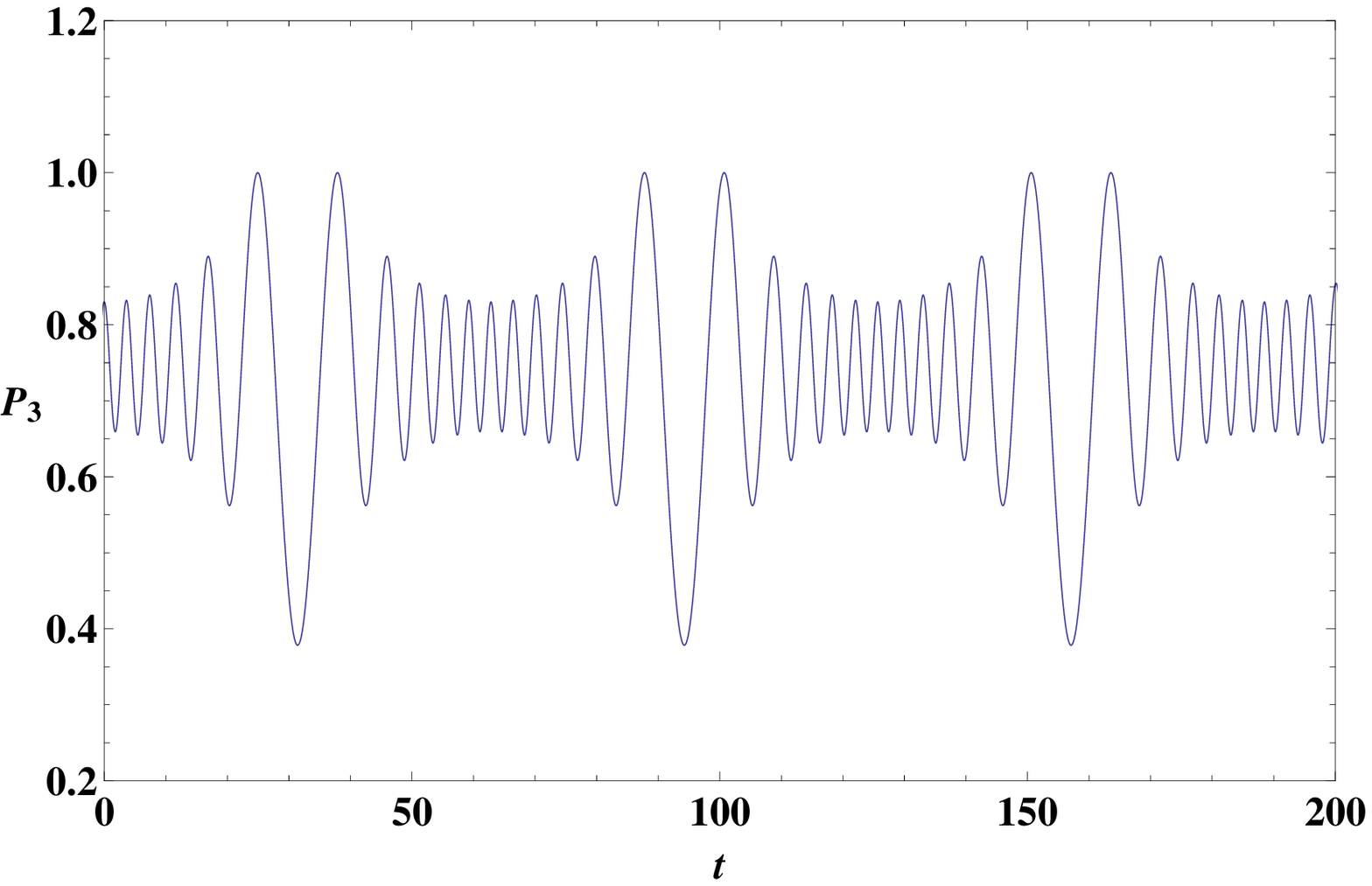}
\caption{{\sl Oscillations of the variable $P_3$ from Eq. (\ref{p3n}) using the same parameters of Fig. \ref{f3} together with $P_{3}(0) = 0.87$}.}
\label{f4} \end{figure}

Let us now consider the opposite case of a very fast plasma process, such as that associated with an electron plasma wave, with typical frequencies $\omega'  \geq \omega_p \gg \omega$, where $\omega_p$ is the electron plasma frequency. In this situation we need to numerically solve the full system of Eqs. (\ref{2.7}) and (\ref{2.7b}).  Figure \ref{f5} below shows the flavor neutrino oscillations found for $\omega_0 = 1.0, \left< \omega \right > = 1.1, \omega' = 3.0, \epsilon = 0.6, \sin^{2}\theta_0 = 0.15$ and initial conditions $P_{1}(0) = P_{2}(0) = 0, P_{3}(0) = 1$. We see the distortion of the flavor population envelope profiles, due to the fast oscillations of the plasma background.

\begin{figure}
\centering
\includegraphics[width=8.5cm]{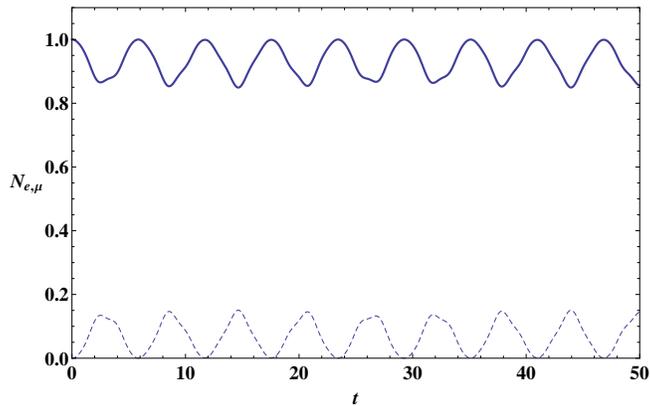}
\caption{{\sl Flavor neutrino oscillations from the direct numerical integration of (\ref{2.7}) and (\ref{2.7b}). Parameters: $\omega_0 = 1.0, \left< \omega \right > = 1.1, \omega' = 3, \epsilon = 0.6, \sin^{2}\theta_0 = 0.15$. Initial conditions: $P_{1}(0) = P_{2}(0) = 0, P_{3}(0)=1$. Line: electron neutrino population $N_e$. Dashed: muon neutrino population $N_\mu$}.}
\label{f5} \end{figure}

\section{Generalized Hamiltonian formulation} 

The dynamical system (\ref{2.7b}) and (\ref{2.8}) can be cast in an generalized Hamiltonian form. This is very useful because it reveals the internal symmetries and invariants of the system. The Hamiltonian formalism has been generalized in several directions since the time of Dirac \cite{dirac}. The importance of generalized Hamiltonians to fluids and plasmas has been detailed in the review by \cite{morrison}. Here we are more concerned with the Nambu generalization to the case of a three-dimensional phase space \cite{nambu}. An extension of Nambu mechanics (which is associated to the $so(3)$ algebra) to arbitrary algebras could also have been considered \cite{morrison2}. 

In order to discuss the Hamiltonian structure of our flavor polarization equations, let us first apply the time re-parametrization (\ref{t}) to obtain
\begin{equation}
\label{r}
\frac{dP_1}{dT} = - P_2 \cos 2\theta \,, \quad  \frac{dP_2}{dT} = P_1 \cos 2\theta - P_3 \sin 2\theta \,, \quad \frac{dP_3}{dT} =  P_2 \sin 2\theta \,. \end{equation}
Second, consider the definition of N-dimensional generalized Hamiltonian systems \cite{morrison,ferhaas}, 
\begin{equation} 
\label{ham}
\frac{dP_i}{dT} = [P_{i},{\cal H}] = \sum_{j=1}^{N} {\cal J}_{ij} \frac{\partial{\cal H}}{\partial P_j} \,,\quad i = 1,\dots,N,
\end{equation}
in terms of local phase-space variables $P_i, i = 1,\dots,N$. Here $[,]$ denotes a generalized Poisson bracket such that 
\begin{equation}
[{\cal A},{\cal B}] = \sum_{i,j=1}^{N}\frac{\partial{\cal A}}{\partial P_i}{\cal J}_{ij}\frac{\partial{\cal B}}{\partial P_j} \,,
\end{equation}
for arbitrary phase-space functions ${\cal A}, {\cal B}$. In (\ref{ham}) the cosymplectic tensor with matrix elements ${\cal J}_{ij}$ should satisfy: (a) ${\cal J}_{ij} = - {\cal J}_{ji}$ (anti-symmetry); (b) $\sum_{l=1}^{N}({\cal J}_{il}\,\partial{\cal J}_{jk}/\partial P_l + {\cal J}_{jl}\,\partial{\cal J}_{ki}/\partial P_l + {\cal J}_{kl}\,\partial{\cal J}_{ij}/\partial P_l) = 0$ (Jacobi identity). Finally, in (\ref{ham}) the function ${\cal H}$ is the Hamiltonian, which is a constant of motion when not explicitly time-dependent. A dynamical system endowed with a generalized Poisson bracket and a Hamiltonian as defined above corresponds to a flow in a Poisson manifold. 

In the present case,  (\ref{r}) is an explicitly time-dependent (through $\theta = \theta(T)$) three-dimensional generalized Hamiltonian system with the Hamiltonian
\begin{equation}
{\cal H} = \frac{1}{2}(P_{1}^2 + P_{2}^2 + P_{3}^2) \,,
\end{equation}
which is clearly a first integral due to the conservancy of the polarization vector modulus, and with 
\begin{equation}
{\cal J}_{ij} =  \left( \begin{array}{ccc}
0 & - \cos 2 \theta & 0 \\
\cos 2 \theta & 0 & - \sin 2 \theta \\
0 & \sin 2 \theta & 0 \end{array} \right) \quad .
\end{equation}

For a given cosymplectic tensor, one can look for a privileged Casimir function ${\cal C}$ which Poisson-bracket commutes with any other phase-space function, so that $\sum_{j=1}^N {\cal J}_{ij}\partial{\cal C}/\partial P_j \equiv 0$. Presently we derive
\begin{equation}
{\cal C} = P_1 \sin 2\theta + P_3 \cos 2\theta \,, 
\end{equation}
which also allows to set the dynamics into the Nambu mechanics \cite{nambu} form 
\begin{equation}
\label{nam}
\frac{d{\bf P}}{dT} = \nabla_{\bf P}{\cal C} \times \nabla_{\bf P}{\cal H} \,,
\end{equation}
where $\nabla_{\bf P} = (\partial/\partial P_{1},\partial/\partial P_{2},\partial/\partial P_{3})$. From Eq. (\ref{nam}) it is apparent that the orbits are found from the intersection between the level surfaces of the Hamiltonian and the Casimir. However, due to the explicit time-dependence ${\cal C}$ is not a constant of motion, 
\begin{equation}
\frac{d{\cal C}}{dT} = - \frac{d\ln\omega}{dT} \frac{dP_2}{dT} \tan 2\theta \,.
\end{equation}
We see that the plasma oscillations induce a non-constant frequency, and hence a time-variance of the Casimir and of the corresponding level surfaces, which are planes in this case.

\section{Influence of a broad turbulence} 

Another important physical situation is that of a plasma with a broad band high frequency turbulent spectrum. In this case, we can use a statistical approach, by taking the average of Eqs. (\ref{2.7}) and (\ref{2.7b}) over a time interval larger than the average flavor oscillating period $1/ \left< \omega \right>$, but much larger than the typical turbulence period $1/ \omega'$. We can then write
\begin{equation}
\frac{d}{d t} \left< P_2 \right> = - a_0 \left< P_3 \right> + \left<  b \right> \left< P_1 \right> + \left<  \tilde b \tilde P_1 \right> ,
\label{4.1} \end{equation}
and
\begin{equation}
\frac{d}{d t} \left< P_1 \right> = - \left< b \right> \left< P_2 \right> - \left< \tilde b \tilde P_2 \right> \; , \quad \frac{d}{d t} \left< P_3 \right>= a_0 \left< P_2 \right> .
\label{4.1b} \end{equation}
Here the symbol $\left<  \right>$ represent the time average  over fast oscillations and tilted variables denote fluctuating parts e.g. $\tilde{P}_j = P_j - \left< P_j \right>$, and where we have defined the quantities
\begin{equation}
a_0 = \omega \sin 2 \theta = \omega_0 \sin 2 \theta_0 \; , \quad b = \omega \cos 2 \theta .
\label{4.2} \end{equation}
On the other hand the fluctuating parts of the polarization vector components have to satisfy the equations
\begin{equation}
\frac{d \tilde P_2}{d t} = - a_0 \tilde P_3 + \left< b \right> \tilde P_1  + \tilde b \left< P_1 \right> ,
\label{4.3} \end{equation}
and
\begin{equation}
\frac{d \tilde P_1}{d t} = - \left< b \right> \tilde P_2 - \tilde b \left< P_2 \right> \; , \quad \frac{d \tilde P_3}{d t} = a_0 \tilde P_2 ,
\label{4.3b} \end{equation}
where the correlations $\left< \tilde b \tilde P_1 \right> - \tilde b \tilde P_1$ and $\left< \tilde b \tilde P_2 \right> - \tilde b \tilde P_2$ were disregarded in view of a weak turbulence assumption. 

We can see from Eqs.(\ref{4.1})-(\ref{4.1b}) that the plasma fluctuations introduce a non-zero contribution to the slow time evolution of ${\bf P}$, due to the average term $\left< b \right>$, which is different from $\left< \omega \right> \left<  \cos 2 \theta \right>$. On the other hand, Eqs.(\ref{4.3})-(\ref{4.3b}) show that these turbulent plasma fluctuations also induce fluctuations in the polarization vector, due to the terms containing $\tilde b$. Coupling between the slow and fast time evolution are also seen in these equations. Let us first consider the fast time scale equations, by using a spectral analysis, through the Fourier decomposition
\begin{equation}
\tilde b = \int \tilde b_\nu e^{- i \nu t} \frac{d \nu}{2 \pi} \; , \quad \tilde P_j = \int \tilde P_{j \nu} e^{- i \nu t} \frac{d \nu}{2 \pi} .
\label{4.4} \end{equation}
Replacing in Eqs.(\ref{4.3})-(\ref{4.3b}), we obtain for the Fourier components of $P_j$ the following results
\begin{equation}
\tilde P_{1 \nu} = - \frac{i}{\nu} f_1 \tilde b_\nu \; , \quad \tilde P_{2 \nu} =  \frac{1}{\delta \nu^2} f_2 \tilde b_\nu \; , \quad \tilde P_{3 \nu} = \frac{i}{\nu} f_3 \tilde b_\nu ,
\label{4.5} \end{equation}
where we have used the auxiliary quantities 
\begin{equation}
f_2 = \left< b \right> \left< P_2 \right> + i \nu \left< P_1 \right> \; , \quad \delta \nu^2 = \nu^2 - ( a_0^2 + \left< b \right>^2) = \nu^2 - \left< \omega \right>^2 ,
\label{4.6} \end{equation}
and
\begin{equation}
f_1 = \frac{f_2}{\delta \nu^2}  \left< b \right> +  \left< P_2 \right> \; , \quad f_3 = \frac{f_2}{\delta \nu^2} a_0 .
\label{4.6b} \end{equation}
These results for the fast time quantities can then be used to calculate the fluctuating terms in the slow time scale equations (\ref{4.1})-(\ref{4.1b}). We obtain
\begin{equation}
\left< \tilde b \tilde P_1 \right> = - i  B_3 \left< b \right> - i \left< P_2 \right> B_1 \; , \quad   
\left< \tilde b \tilde P_2 \right> =   B_2 \; , \quad
\left< \tilde b \tilde P_3 \right> = i a_0 B_3.
\label{4.7} \end{equation}
The new quantities $B_j$ appearing in these expressions are functions of the energy content of the plasma turbulence spectrum, and are defined by  
\begin{equation}
B_1 =  \left< \frac{\tilde b^2}{\nu} \right> \; , \quad B_2 =  \left< \frac{f_2 \tilde b^2}{\delta \nu^2} \right> \; , \quad B_3 =  \left< \frac{f_2 \tilde b^2}{\nu \delta \nu^2} \right> ,
\label{4.8} \end{equation}
where the time average operation is equivalent to an integration over the turbulent spectrum, as indicated: $ B_0 \equiv \left< \tilde b^2  \right> = \int | \tilde b_\nu |^2 d \nu / 2 \pi$. For electron plasma turbulence, we can obviously use the estimate: $B_j = B_0 / \omega_p^j$, with $j =1, 2, 3$, where $\omega_p$ is the electron plasma frequency. Replacing these results in equations (\ref{4.1})-(\ref{4.1b}), we finally get
\begin{equation}
\frac{d}{d t} \left< P_2 \right> = - a_0 \left< P_3 \right> + \left<  b \right> \left< P_1 \right>  - i (B_3 \left<  b \right> + B_1 \left< P_2 \right>) ,
\label{4.9} \end{equation}
and
\begin{equation}
\frac{d}{d t} \left< P_1 \right> = - \left< b \right> \left< P_2 \right> - B_2 \; , \quad \frac{d}{d t} \left< P_3 \right>= a_0 \left< P_2 \right> .
\label{4.9b} \end{equation}

A closed equation for the component $\left< P_2 \right>$ can be obtained by taking the time derivative of Eq. (\ref{4.9}), and using Eq. (\ref{4.9b}). If we further assume steady state turbulence we get
\begin{equation}
\left[ \frac{d^2}{d t^2}  + i B_1 \frac{d}{d t}  + \left< \omega \right>^2 \right] \left< P_2 \right> = - B_2  \left<  b \right> \,, 
\label{4.10} \end{equation}
whose solution is 
\begin{equation}
\left< P_2 \right> = - \frac{B_2 \left< b \right>}{\left< \omega \right>^2} + \alpha e^{i\Omega_+ t} + \beta e^{i\Omega_- t} + {\rm c.c.} \,,
\end{equation}
where $\alpha$ and $\beta$ are integration constants and where
\begin{equation}
\Omega_{\pm} = - \frac{B_1}{2} \pm \left(\frac{B_{1}^2}{4} + \left<\omega\right>^2\right)^{1/2} \,.
\end{equation}

We conclude that the effects of plasma turbulence are two-fold: to shift both the frequency oscillation and the equilibrium value of the average coherence $\left< P_2 \right>$. 
We can easily realize that the  shifting frequency is of the order of
\begin{equation}
B_1 \simeq \frac{B_0}{\omega_p} \simeq  \frac{\left< \omega \right>^2}{\omega_p} \left< \Bigl(\frac{\tilde n}{n_0}\Bigr)^2 \right> \,,
\label{4.11} \end{equation}
where $\tilde n$ are the electron density perturbations and $n_0$ the equilibrium density. In what concerns the turbulent force $- B_2  \left<  b \right>$, we can see that it depends on $B_2 \simeq B_{1}/\omega_p$ and $\left< b \right> = \left< \omega \cos 2 \theta \right>$. 

\section{Conclusions}

We have studied the influence of electron plasma oscillations on the neutrino flavor instabilities. We have first shown that slow changes in the plasmas medium, with typical frequencies much smaller than the neutrino flavor frequency, lead to a spectral broadening of the neutrino oscillation process and to beat wave disturbances of these oscillations. On the other hand, a single high frequency plasma oscillation will introduce high frequency flavor processes, which can eventually become unstable. In this case, quantum coherence can increase, as imposed by the oscillation of the background medium.  In spite of the explicit time-dependence, the exact conservation law of $|{\bf P}|$ can be used to cast the model in a generalized Hamiltonian form. 

We have also studied the influence of a broadband high frequency turbulent spectrum. In this case, turbulence can be described as a stochastic process and a statistical approach similar to that used in the Langevin model can be used. Equations for the long time evolution of the flavor polarization vector were derived. This evolution is intimately dependent on the fast time evolution processes, and as a result,  shifting frequency and force terms occur due to the presence of turbulence. 

These results definitely show that plasma oscillations can be intimately linked with the quantum coherent processes associated with flavor oscillations. The amplitude of the plasma oscillations can be, on the other hand, influenced by these flavor oscillations, and a more complete description of the neutrino plasma coupling needs to address this mutual influence, which will be considered in a forthcoming publication. Our results can be used for qualitative and semi-quantitative analysis, because they only concern a two-flavor model. In the case of a more rigorous three flavor model, we would be faced with a non-autonomous system of 8 first-order ordinary differential equations, hardly amenable to meaningful approximate methods (see e.g. \cite{hollenberg}.

\begin{acknowledgments}

One of us (JTM) would like to thank the financial support of CAPES, and the hospitality of Professor Ricardo Galv\~ao during his stay at the Institute of Physics of the University of S\~ao Paulo. FH acknowledges the financial support by CNPq (Conselho Nacional de Desenvolvimento Cient\'{\i}fico e Tecnol\'ogico). 

\end{acknowledgments}

\end{document}